\documentclass[a4]{article}

\usepackage[lofdepth,lotdepth]{subfig}
\usepackage{graphicx}
\usepackage{algorithm}
\usepackage{algpseudocode}
\usepackage{cite}
\usepackage{amssymb}
\usepackage{gensymb}
\usepackage{hyperref}
\usepackage{multicol}

\usepackage[latin1]{inputenc}
\usepackage{tikz}
\usetikzlibrary{shapes,arrows,positioning,calc}

\newcommand{\bfm}[1]{\textbf{#1}}

\newcommand{\R}[1]{{\color{black}#1}}

\newcommand{\dif}{{\rm d}}

\newcommand{\vJ}{\bfm{J}}

\newcommand{\vE}{\bfm{E}}
\newcommand{\vB}{\bfm{B}}

\newcommand{\vG}{\bfm{G}}

\newcommand{\vr}{\bfm{r}}

\newcommand{\vnh}{\hat\bfm{n}}
\newcommand{\vxh}{\hat\bfm{x}}
\newcommand{\vyh}{\hat\bfm{y}}
\newcommand{\vzh}{\hat\bfm{z}}

\newcommand{\ten}[1]{\overline{\overline{#1}}}

\tikzstyle{startstop} = [rectangle, rounded corners, minimum width=3cm, minimum height=1cm,text centered, draw=black, fill=red!30]
\tikzstyle{io} = [trapezium, trapezium left angle=70, trapezium right angle=110, minimum width=3cm, minimum height=1cm, text centered, draw=black, fill=blue!30]
\tikzstyle{process} = [rectangle, minimum width=3cm, minimum height=1cm, align=left, draw=black, fill=orange!30]
\tikzstyle{empty} = [rectangle, align=left]
\tikzstyle{decision} = [diamond, minimum width=3cm, minimum height=1cm, align=center, draw=black, fill=green!30]
\tikzstyle{arrow} = [thick,->,>=stealth]

\begin{document}
%
\title{Electro-thermal modelling by novel variational methods: racetrack coil in short-circuit\footnote{{\bf Please, note that copyright applies for IEEE}. Accepted version of the article published as E Pardo and A Dadhich, IEEE Trans. Appl. Supercond., vol. 33, no. 5, a.n. 5201606, year 2023. Related DOI: \url{https://doi.org/10.1109/TASC.2023.3252492}.}}

\author{Enric Pardo\footnote{Corresponding author: enric.pardo@savba.sk}, Anang Dadhich \\
\normalsize{Institute of Electrical Engineering, Slovak Academy of Sciences,} \\
\normalsize{Bratislava, Slovakia.}
}




\maketitle

\begin{abstract}
The design of superconducting applications containing windings of superconducting wires or tapes requires electro-thermal quench modelling. In this article, we present a reliable numerical method based on a variational principle and we benchmark it to a conventional finite difference method that we implemented in C++. As benchmark problem, we consider a racetrack coil made of REBCO superconducting tape under short circuit, approximated as a DC voltage that appears at the initial time. Results show that both models agree with each other and analytical limits. Since both models take screening currents into account, they are promising for the design of magnets (especially fast-ramp magnets) and power applications, such as the stator windings of superconducting motors or generators.
\end{abstract}



\section{Introduction}

\begin{figure}[tbp]
\centering
{\includegraphics[trim=10 20 20 12,clip,width=4.3 cm]{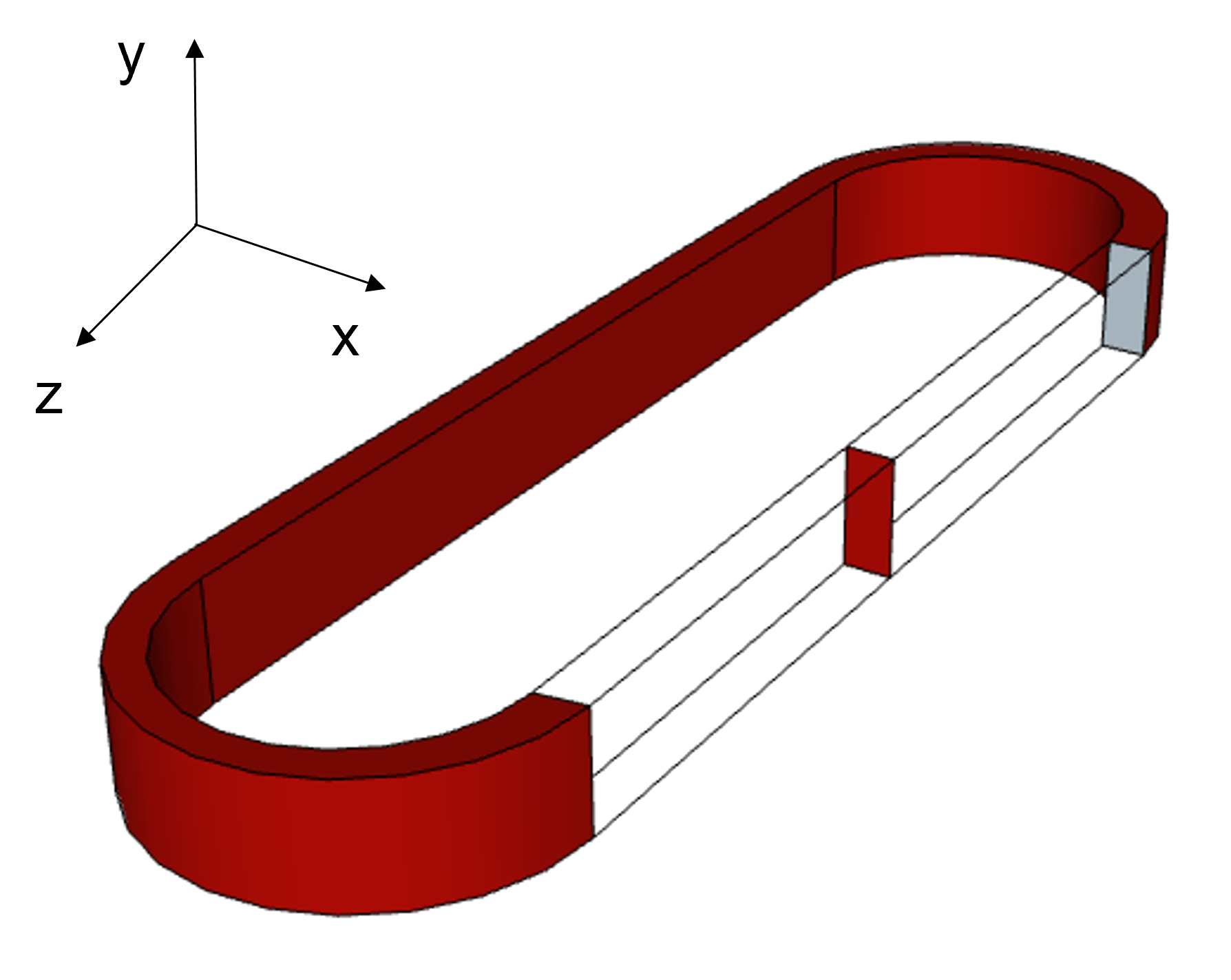}}
{\includegraphics[trim=600 220 300 50,clip,width=4.3 cm]{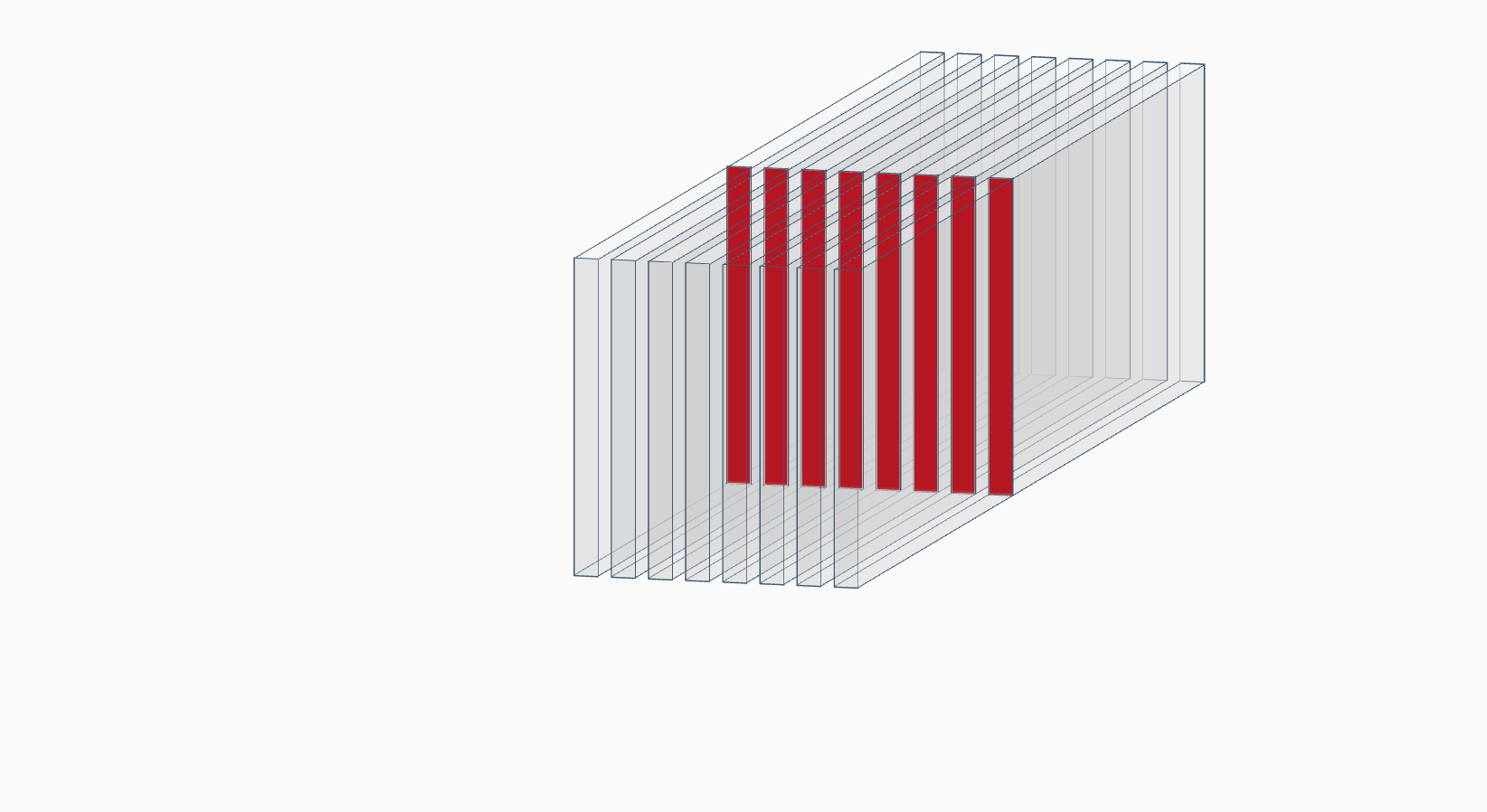}}\\
{\includegraphics[trim=0 0 0 0,clip,width=8 cm]{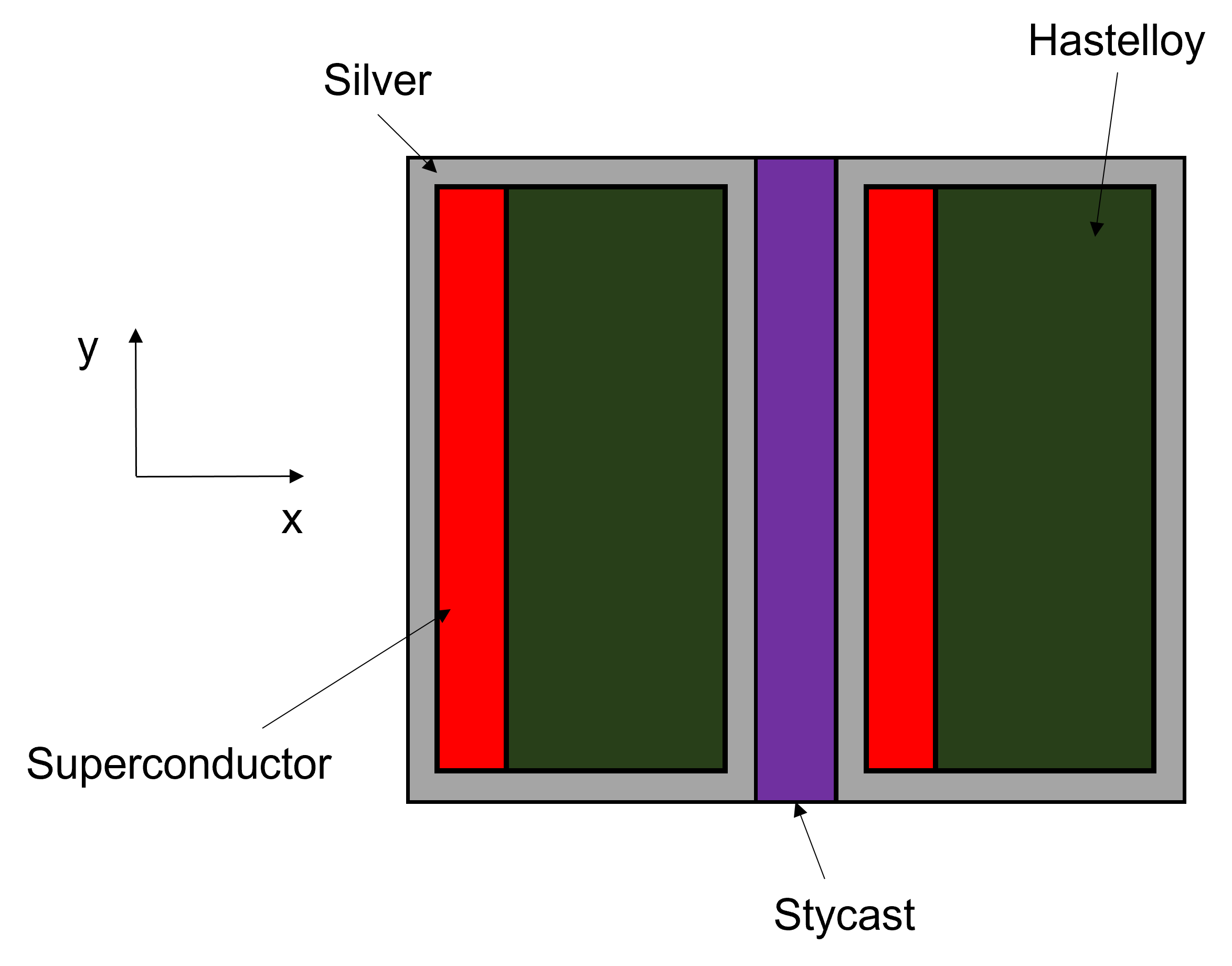}}
\caption{(Top) Sketch of the modelled REBCO ractrack coil and its cross-section. (Bottom) Details of the tape cross-section and electrical \R{insulation}.}
\label{fsketch}
\end{figure}

The design of superconducting applications containing windings of superconducting wires or tapes requires electro-thermal modelling. Indeed, electro-thermal quench is a main concern for magnets \cite{hahnS2019Nat, baiH2020IESa, awaji2S021IES, badelA2021IES, fazilleauP2018SST, lecrevisseT2022SSTa, yoonSK2016SST, bhattaraiKR2020SST, shaoL2021Ele} and the stator or rotor windings of motors and generators \cite{bergenA2019SST}.

Although there has been extensive developments in electro-thermal modeling in magnets \cite{chanWK2012IES, breschiM2017IES, noguchiS2019IES, markiewiczWD2019SST, badelA2019SST, bhattaraiKR2020SST, gavrilinAV2021IES, niuM2021AMM, genotC2022IES} and fault-current limiters or single tapes \cite{royF2008IES, lacroixC2014SST, bonnardCH2016SST, rivaN2020SST, gomoryF2022IES}, these usually neglect screening currents. The reason is that taking screening currents into account is highly time consuming \R{and} of little relevance for currents well above the critical current. However, heating from screening currents is key for windings working at high frequencies, such as stator windings in motors, or magnets under high sweep rates or emergency shut-downs. In addition, electro-thermal effects could influence the screening current induced field, which should also be taken into account in magnet design.

Then, there is a \R{need to develop} fast and robust numerical methods to predict the electro-thermal behaviour of superconducting windings. In this article, we present a novel electro-thermal method based on variational \R{principles} and we benchmark it to a finite difference method coupled with an electromagnetic modeling tool that we implemented in C++. Both methods take screening currents into account. As an example, we model a REBCO racetrack coil under a shortcut, considered as a sharp rise of voltage up to a DC value. This study is useful not only as a benchmark of the models but also to understand the electro-thermal behavior of racetrack coils under these conditions.


\section{Configuration}

\begin{figure}[tbp]
\centering
{\includegraphics[trim=0 0 0 0,clip,width=8 cm]{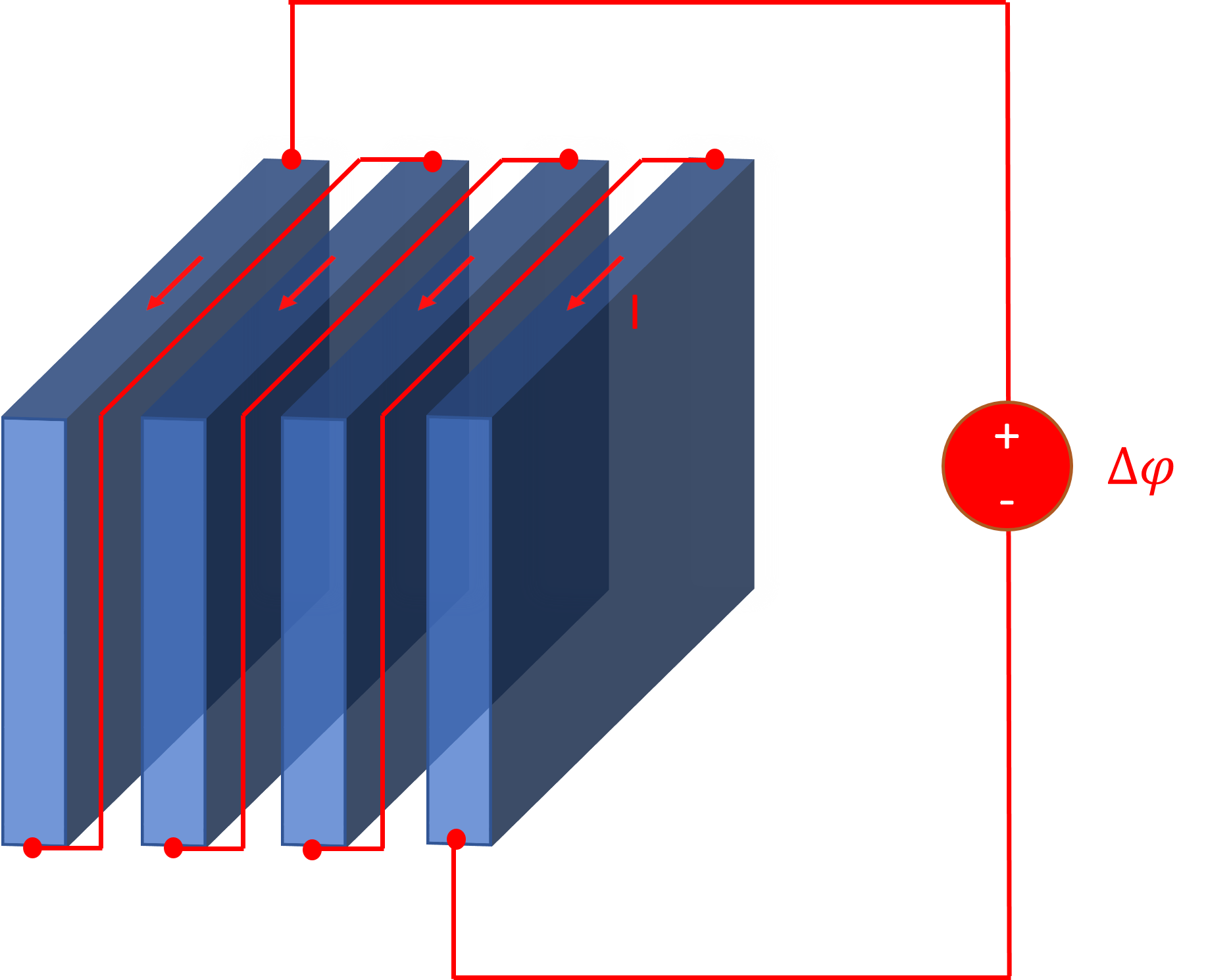}}
\caption{Infinitely long approximation of the straight part of the coil under an applied voltage, $\Delta\phi$. \R{Only 4 of the 8 turns of the coil are shown.}}
\label{fsketch}
\end{figure}

In this article, we analyze the case of a racetrack coil in short circuit. Initially, the coil is submitted to zero voltage (and current) and \R{immediately} afterwards the voltage, $\Delta \phi$, suddently switches to a given DC value. This configuration is of practical importance, specially when the DC voltage is large enough to cause electro-thermal quench. We also use this example to \R{benchmark} between the two presented numerical methods (section \ref{smethod}), as well as between the methods and analytical limits.

In particular, we consider a racetrack coil of 8 turns of 4 mm wide REBCO superconducting tapes (figure \ref{fsketch}), where the straight part is much longer than the semi-circular ends. We also assume a bore width of 1 m. The tapes are 114 $\mu$m thick with 2 $\mu$m of silver stabilization on all sides, 100 $\mu$m thick Hastelloy substrate, and 10 $\mu$m of REBCO superconducting layer. The turns are \R{insulated} by 50 $\mu$m of Stycast (figure \ref{fsketch}).

We take the non-linear heat convection factor, $h$, for liquid nitrogen at 77 K from \cite{royF2008IES}. For simplicity, we take constant material properties. Heat conductivities: 9, 400, 7, 0.8 W/Km for superconductor, silver, Hastelloy, and \R{Stycast}; heat capacities 162, 2470, 3700, 317 kJ/Km$^3$ for superconductor, silver, Hastelloy, and \R{Stycast}; resistivity (or normal-state resistivity) 3$\cdot 10^{-7}$, $10^{-8}$, 1.2$\cdot 10^{-6}$, $\infty$ $\Omega\cdot$m for superconductor, silver, Hastelloy, and \R{Stycast}; and constant superconductor critical current density, corresponding to a critical current of 150 A. The normal transition temperature is taken as 92 K.


\section{Modeling Method}
\label{smethod}

Here, we introduce the Minimum Electro-Thermal Entropy Production (METEP) method and we benchmark it with a finite difference method. In the examples of this work, we assume that the straight part of the racetrack coil is much longer than the semi-circular segments. Thus, we can take the infinitely long approximation to reduce the mathematical problem to the 2D cross-section (figure \ref{fsketch}). However, the METEP method is valid for any 3D shape.


\subsection{Minimum Electro Thermal Entropy Production method}

Generally, the relation between the electric field, $\vE$, and current density, $\vJ$, depends on the magnetic induction and temperature, $\vE(\vJ,\vB,T)$. The heat conductivity tensor, $\ten{k}$, and capacity, $C_v$, also depend on temperature, $\ten{k}(T)$, $C_v(T)$. The method presented here is valid for any of these relations, although the examples in this work take certain simplifications into account (section \ref{sproperties}).

Our method \R{solves} the thermal diffusion equation for the temperature ($T$), which can be written as
\begin{equation} \label{Tdiff}
p=\frac{\partial U_T(T)}{\partial t}+\nabla\cdot\vG(\nabla T)
\end{equation}
with
\begin{equation}
\vG(\nabla T)=-\ten{k}\nabla T, \quad U_T(T)\equiv\int_0^T\dif T'C_v(T'),
\end{equation}
where $p$ is the power density dissipation ($p=\vE\cdot\vJ$ for Joule heat), $\vG$ is the heat flux density, and $t$ is the time. By now, we assume $\ten{k}$ as temperature independent, but later we take $\ten{k}(T)$ into account. By discretizing the time evolution into finite time steps, (\ref{Tdiff}) becomes
\begin{equation} \label{Tdiff_fin}
\R{p=\frac{U_T(t)-U_T(t-\Delta t)}{\Delta t}+\nabla\cdot\vG(\nabla T)}.
\end{equation}

Following the same steps as in \cite{pardoE2017JCP} for the electro-magnetic model, we find that solving (\ref{Tdiff_fin}) for a known initial temperature, $T_0\equiv T(t-\Delta t)$, is the same as minimizing the functional
\begin{eqnarray}
&& F[T]=\int_V\dif V \Big\{ [h(T)-U_T(T_0)T]/\Delta t \nonumber \\
&&\R{+ \frac{1}{2}\nabla T \ten{k} \nabla T} - q(\vJ,\vB,T) \Big\} \label{FT}
\end{eqnarray}
with
\begin{equation}
h(T)\equiv\int_0^T\dif T'U_T(T'), \quad q(\vJ,\vB,T)\equiv\int_0^T\dif T'\vJ\cdot\vE(\vJ,\vB,T),
\end{equation}
where $\vJ\cdot\vE=p$. Indeed, (\ref{Tdiff_fin}) is the Euler equation of (\ref{FT}). In addition, the second variation of the functional is always positive because $C_v$ is positive for any $T$ and $\ten{k}$ is positive-definite thanks to the second law of thermodynamics. As a consequence, there \R{exists} a single minimum of the functional, and hence the solution is unique for any given time step. We can calculate the whole time evolution, if we know the initial temperature field, $\vJ(t)$, and $\vB(t)$.

Next, we discretize our object into elements where $T$ is assumed uniform in the volume and the $\nabla T$ component perpendicular to the faces is uniform there. For any element, $i$, and each of its faces, labeled as $\nu$, $\nabla T$ is obtained as
\begin{equation}
(\nabla T)_{i\nu}=\frac{(T_i-T_{m(i\nu)})}{\|\vr_i-\vr_{m(i\nu)}\|^2} (\vr_i-\vr_{m(i\nu)}),
\end{equation}
where $T_i$ is the temperature in element $i$, and $m(i\nu)$ is the index of the element next to surface $\nu$ of element $i$. Next, we assume brick elements of size $\Delta x_i\times\Delta y_i\times\Delta z_i$, although the functional of (\ref{FT}) is valid for any mesh. \R{For this mesh, the heat flow across each surface is $G_{i\nu}=-(1/2)(\ten{k}_{i}+\ten{k}_{m(iv)})(\nabla T)_{i\nu}$, where $\ten{k}_i$ is the heat conductivity tensor at the center of element $i$.} After dividing the body into $N$ brick elements, the functional becomes
\begin{eqnarray} \label{FTdis}
F({\tilde T})\approx \sum_i^N V_i\Big\{[h(T_i)-U(T_{0i})T_i]/\Delta t - q(\vJ_i,\vB_i,T_i)+f_{i} \Big\} 
\end{eqnarray}
with
\begin{equation} \label{fi}
\R{f_{i}\equiv\frac{1}{4}\sum_{\nu=1}^{n_{\rm face}}(\nabla T)_{i\nu}\frac{1}{2} (\ten{k}_{i} + \ten{k}_{m(i\nu)} ) (\nabla T)_{i\nu},}
\end{equation}
where $\tilde T$ is a data vector containing $T_i$ for all elements $i$, and $n_{\rm face}$ is the number of faces per element (6 for brick elements). Above, $f_i$ contains a factor 1/4 instead of 1/2 because each surface belongs to two elements. For orthotropic materials, $\ten{k}$ is diagonal with components $k_x$, $k_y$, $k_z$ corresponding to directions $x$, $y$, $z$. Then, $f_i$ in (\ref{fi}) can be simplified into
\begin{equation}
\R{f_i=\frac{1}{8}\sum_{\nu=1}^{n_{\rm face}}(\nabla T)_{i\nu}^2 (k_{\nu,i} + k_{\nu,m(i\nu)}) .}
\end{equation}
\R{When surface $\nu$ is perpendicular to the $x$ axis, $k_{\nu,i}$ becomes $k_{x,i}$, and so on for surfaces perpendicular to the $y$ and $z$ axis.}

In order to take the heat exchange with a cryogenic liquid into account, we consider an additional layer of volume elements on the surface, where the temperature is that of the liquid, $T_l$. Afterwards, we use an effective thermal conductivity that results in the same heat flux density. For a body immersed in a liquid, the heat flux density is
\begin{equation} \label{Gh}
\vG = h(T-T_l)\cdot(T-T_l)\cdot \vnh,
\end{equation}
where $h$ is the (non-linear) convection coefficient, $T$ is the temperature at the object surface, and $\vnh$ is the normal unit vector outwards the surface. For a surface perpendicular to $\vxh$, the heat flux density due to the equivalent conductivity, $k_{sx}$, is $\vG=k_{sx}\frac{(T-T_l)}{\Delta x}\vxh$. Comparing to (\ref{Gh}), the equivalent conductivity is $k_{sx}=h(T-T_l)\Delta x$. Similarly, the equivalent conductivities for the sufaces perpendicular to $\vyh$ and $\vzh$ are $k_{sy}=h(T-T_l)\Delta y$ and $k_{sz}=h(T-T_l)\Delta z$.

We find the temperature \R{in} all elements, $\tilde T$, by minimizing the functional (\ref{FTdis}) in mainly the same way as in \cite{pardoE2015SST}. We find the element $j_+$ where changing the temperature by $\delta_T=+d$, with $d>0$, causes the smallest change in the functional. We do the same for $\delta_T=-d$, finding the minimizing element $j_-$. If one of these changes in functional is negative, we apply the corresponding change in temperature (either $\delta_T=+d$ at $j_+$ or $\delta_T=-d$ at $j_-$, according to which change in temperature reduces the functional the most). Then, continue the same steps until any change of temperature of $\delta_T=\pm d$ does not decrease the functional. For temperature dependent $\ten{k}$, we later update \R{$k_{x,i}$, $k_{y,i}$, $k_{z,i}$} in all elements, according to the recently found $\tilde T$ and iterate until the difference in temperature is below $|\delta_T|$ for any element. 

For the evaluation of the change of functional due to a change $\delta_T$ at element $j$, $\delta F_j(\delta_T)$, it is convenient to use the following approximations in order reduce the number of numerical integrals
\begin{eqnarray}
q(T+\delta_T)-q(T)\approx \delta_T \vJ\cdot[ \vE(T+\delta_T)+\vE(T) ]/2 \\
h(T+\delta_T)-h(T)\approx \delta_T [U_T(T+\delta_T)+U_T(T)]/2.
\end{eqnarray}
\R{For} $q$, we did not write the dependence on $\vJ$ and $\vB$ for notation simplicity.

In general, $\vJ$ and $\vB$ are not inputs to the problem, but solutions of electromagnetic computations. We performed full electro-thermal modelling that take screening currents into account as follows. At a given $T$ field, we find $\vJ$ (and $\vB$) by the electromagnetic model. Then, we use these $\vJ$ and $\vB$ to find $T$ by METEP. Later, we iterate until the difference in both $\vJ$ and $T$ are below certain tolerances. In this article, we use the electro-magnetic method in section \ref{sEM}.


\subsection{Finite difference method}

In order to benchmark METEP, we also implemented a finite difference method in C++ for the thermal model \cite{powellA2002prp}. In particular, we use the explicit finite difference method. For this case, the electromagnetic model outputs the Joule power density $p$, which is the input for the thermal model. In order to achieve stable temperature distributions from the finite difference method, the time step should follow $\Delta t\le {C_v\min(\Delta x^2,\Delta y^2)}/{2k}$. Since the finite difference method requires very short time steps, we enable many time steps of finite difference calculations within the time interval of the electromagnetic model. The final solution of $\vJ$ and $T$ \R{at} the end of each electromagnetic time step is achieved by iterating the electromagnetic and thermal model until the difference is below a certain tolerance. For simplicity, here we assume temperature-independent $C_v$ and $k$ \R{in} the sample volume.


\subsection{Electromagnetic model} \label{sEM}

For the electromagnetic model, we use the Minimum Electro-Magnetic Entropy Production (MEMEP) method \cite{pardoE2015SST, pardoE2017JCP, ghabeliA2021SST} but with a voltage, $\Delta\phi$, as input. For the assumed infinitely long geometry of figure \ref{fsketch} the current density follows $\vJ=J\vzh$. The solution of $J$ minimizes the functional
\begin{equation}
F_J=l\int_S\dif S\left [ \Delta J\frac{\Delta A_J}{2\Delta t} + \Delta J\frac{\Delta A_a}{\Delta t} + U(J) \right ] - I\Delta\phi
\end{equation}
with $U(J)=\int_0^J\dif J'E(J')$, where $l$ is the racetrack length, $S$ is the whole cross-section, $I$ is the net current at each turn obtained by the contribution of all elements, $A_J$ and $A_a$ are the vector potential generated by $J$ and external sources, respectively, $\Delta J=J(t)-J(t-\Delta t)$, and $\Delta A_J$ and $\Delta A_a$ are defined similarly as $\Delta J$.


\subsection{Homogenized material properties}
\label{sproperties}

\begin{figure}[tbp]
\centering
{\includegraphics[trim=0 27 0 0,clip,width=9 cm]{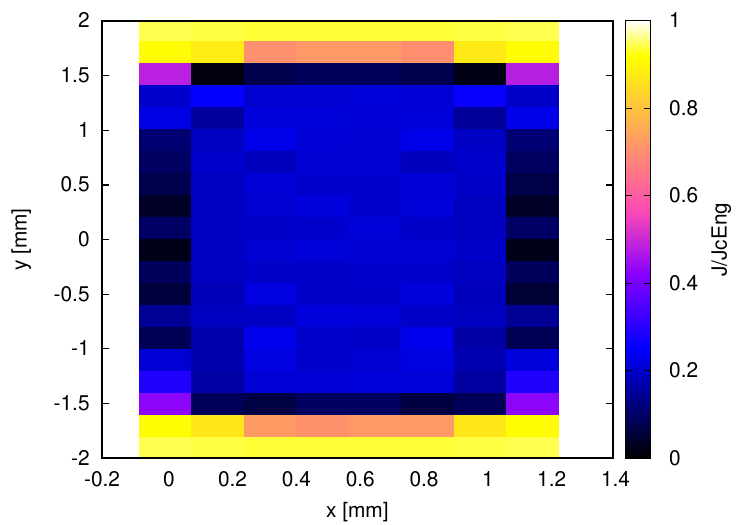}}\\
{\includegraphics[trim=0 0 0 0,clip,width=9 cm]{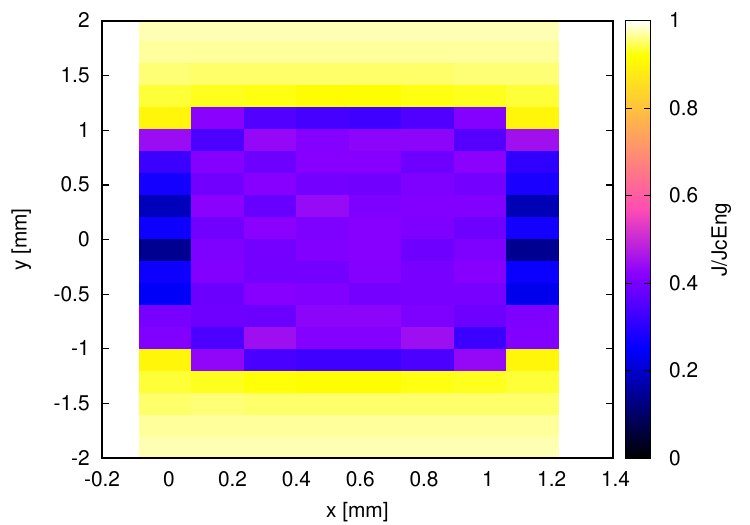}}
\caption{Engineering current density after 0.05 and 0.1 s after switching on a DC voltage per tape length of 0.01 V/m.}
\label{fJmap}
\end{figure}

\begin{figure}[tbp]
\centering
{\includegraphics[trim=0 25 0 0,clip,width=9 cm]{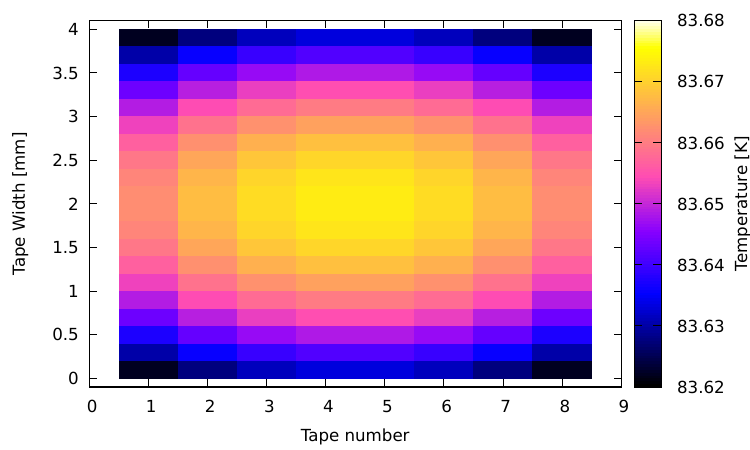}}\\
{\includegraphics[trim=0 0 0 0,clip,width=9 cm]{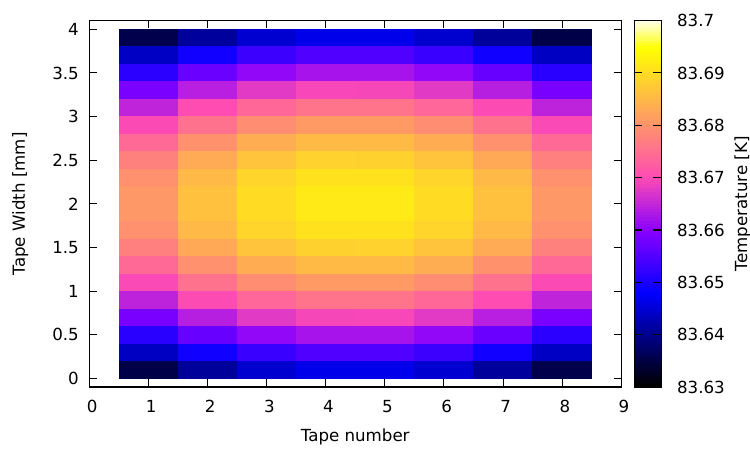}}
\caption{The temperature distribution from METEP (a) and the finite difference method (b) agree. Case for 40 s after switiching on a voltage of 0.01 V/m.}
\label{fTmap}
\end{figure}

\R{For our 2D cross-section, $\vJ$ and $\vE$ follow the $z$ direction, and hence $\vJ=J\hat{\bf z}$ and $\vE=E\hat{\bf z}$.} For the superconducting current, $\vJ_s$, we assume a power-law \R{$E(J_s)$} relation: \R{
\begin{eqnarray}
&& E=E_c(|J_s|/J_c)^n{\rm sign}(J_s) \nonumber \\
&& J_s=J_c(|E|/E_c)^{1/n}{\rm sign}(E) .
\end{eqnarray} }
Next, we consider a homogeneous model for the tape, where there are also normal conducting parts. These contain the metal layers, as well as the normal conducting contribution from the superconductor. All these materials are in parallel, and hence the effective normal resistivity of the tape is 
\begin{equation}
\rho_n=\left ( \sum_{m=1}^{n_m} \frac{S_m}{S\rho_m} \right )^{-1},
\end{equation}
where $S$ is the total tape section, $S_m$ is the section of the material with label $m$, and $n_m$ is the number of different materials. The total current is the sum of the superconductor and normal currents. Then, the engineering current density, $J$, follows \R{$J=J_n+J_sS_s/S$}, being $S_s$ the superconductor section. Therefore
\begin{equation}
J=E/\rho_n + \R{J_{ce}}(|E|/E_c)^{1/n}{\rm sign(E)} \equiv g(E),
\end{equation}
where \R{$J_{ce}$} is the enginering $J_c$, \R{$J_{ce}=J_cS_s/S$}. Then, the $E(J)$ relation of the homogeneous material is
\begin{equation}
\R{E(J)=g^{-1}(J)},
\end{equation}
where \R{$g^{-1}(J)$} is the inverse function of \R{$g(E)$}. The $T$ and $\vB$ dependence of $\vE$ for a given $\vJ$ are introduced via $T$ and $\vB$ dependencies of $J_c$ and $n$. In the results in this article, we assume constant $n$, $\vB$-independent $J_c$ and the following $T$ dependence of $J_c$ \R{for $T$ below $T_c$}:
\begin{equation}
J_c(T)=J_c(T_l)\frac{T_c-T}{T_c-T_l},
\end{equation}
where $T_c$ is the normal-state transition temperature. \R{Naturally, $J_c=0$ for $T\ge T_c$.}

For the thermal properties, we assume a homogeneous orthotropic model. Given $C_v(x,y)$, $k_x(x,y)$ and $k_y(x,y)$ due to different materials including the inter-turn isolation, the effective quantities are
\begin{eqnarray}
C_{ve} & = & \frac{1}{S}\int\dif x\dif y C_v(x,y) \\
k_{xe} & = & \frac{1}{D_y}\int\dif y\left [ \frac{1}{D_x}\int\dif x\frac{1}{k_x(x,y)} \right ]^{-1} \\
k_{ye} & = & \frac{1}{D_x}\int\dif x\left [ \frac{1}{D_y}\int\dif y\frac{1}{k_y(x,y)} \right ]^{-1},
\end{eqnarray}
where $D_x$ and $D_y$ are the tape sizes in the $x$ and $y$ directions.


\subsection{Ananlytical limit} \label{s.analytical}

We also benchmark the models with the analytical limit with the following assumptions: infinitely long coil, uniform $T$, uniform normal conducting material, and constant $\rho$, $C_v$, $h$. If we also neglect the coil inductance, the solution of the thermal diffusion equation after switching on the voltage at $t=0$ is
\begin{equation}
T(t)=(T_\infty-T_l)(1-e^{-t/\tau})+T_l
\end{equation}
with $T_\infty=v^2S/(\rho hD)+T_l$ and $\tau=C_vS/(Dh)$, where $v$ is the voltage per unit tape length, $S$ is the coil cross-section, and $D$ is the perimeter of the coil cross-section.


\section{Results and Discussion}

\begin{figure}[tbp]
\centering
{\includegraphics[trim=0 0 0 0,clip,width=9 cm]{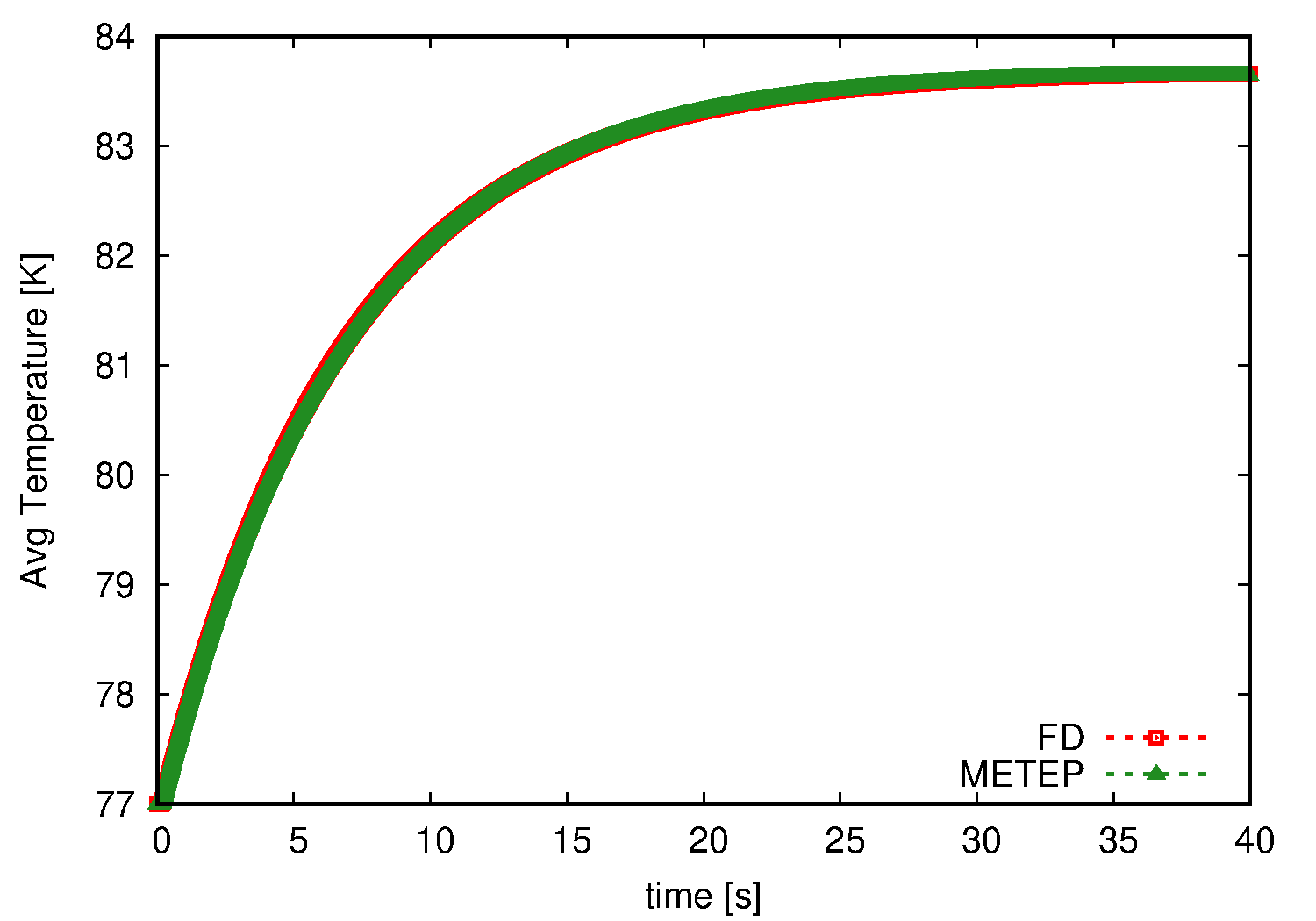}}\\
{\includegraphics[trim=0 0 0 0,clip,width=9 cm]{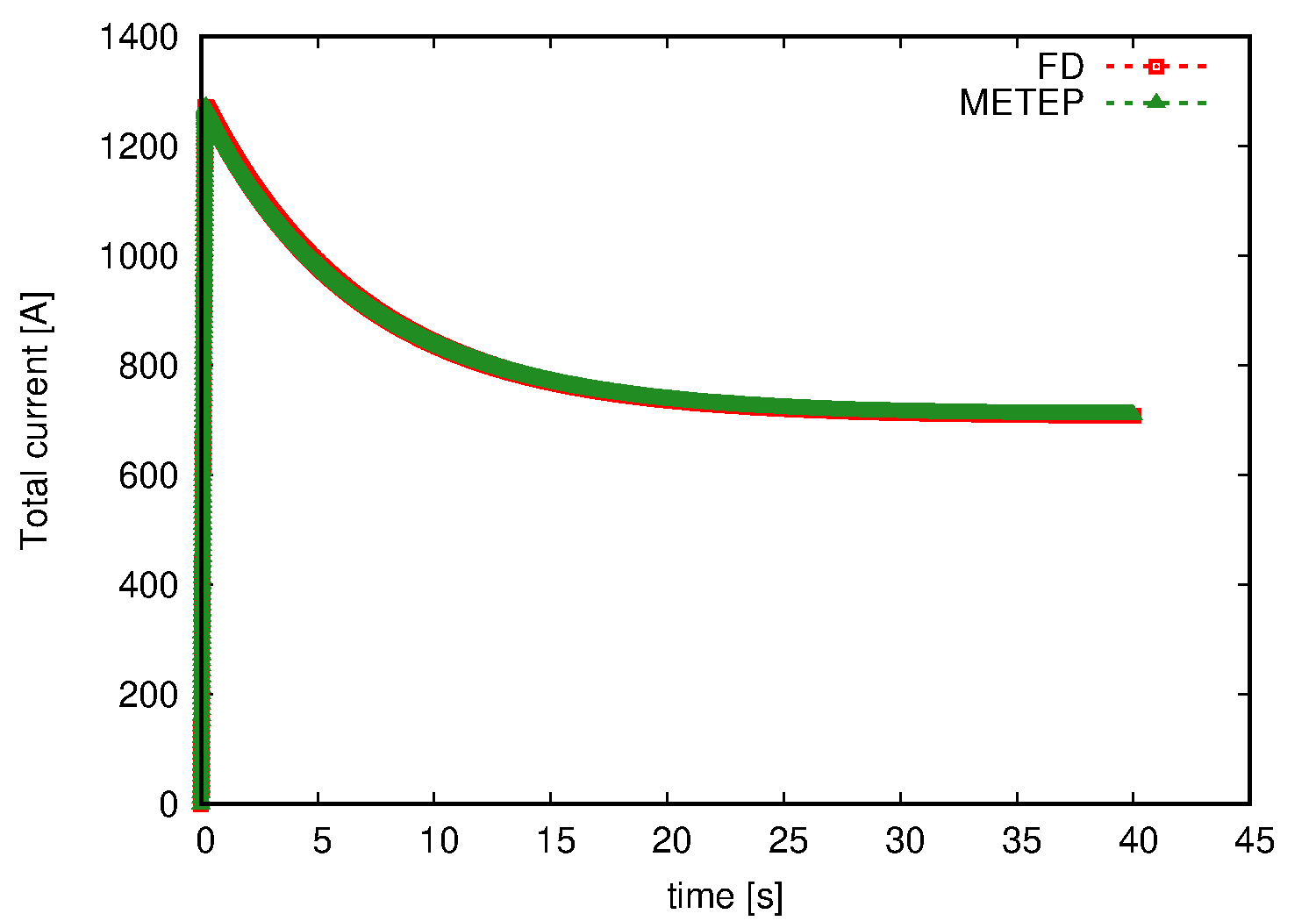}}
\caption{Evolution of the average temperature in the coil section (top) and the total current in A$\times$turns for an input voltage of 0.01 V/m. Calculations from METEP and finite differences (FD), respectively.}
\label{fTlowV}
\end{figure}

\begin{figure}[tbp]
\centering
{\includegraphics[trim=0 0 0 0,clip,width=9 cm]{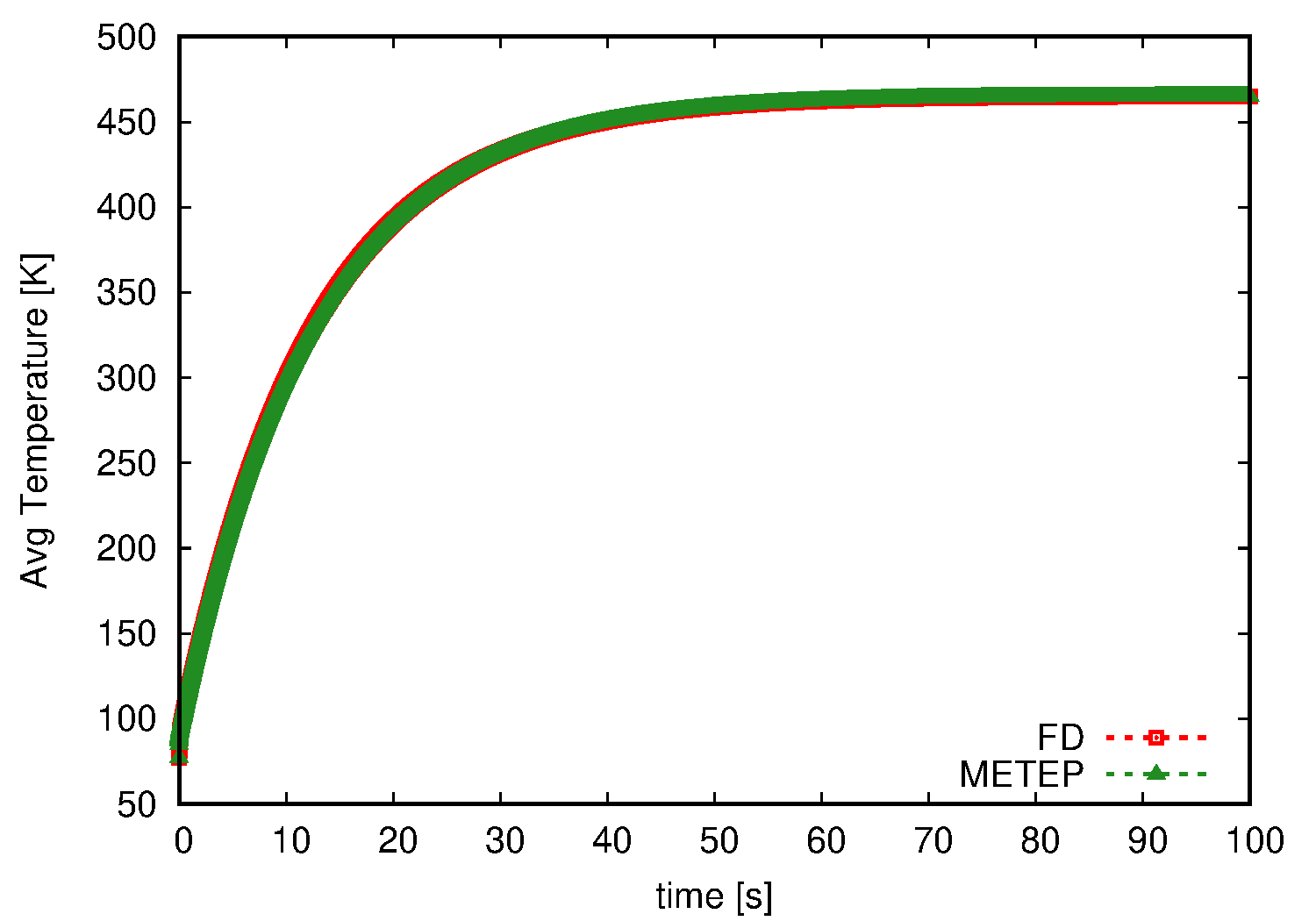}}\\
{\includegraphics[trim=0 0 0 0,clip,width=9 cm]{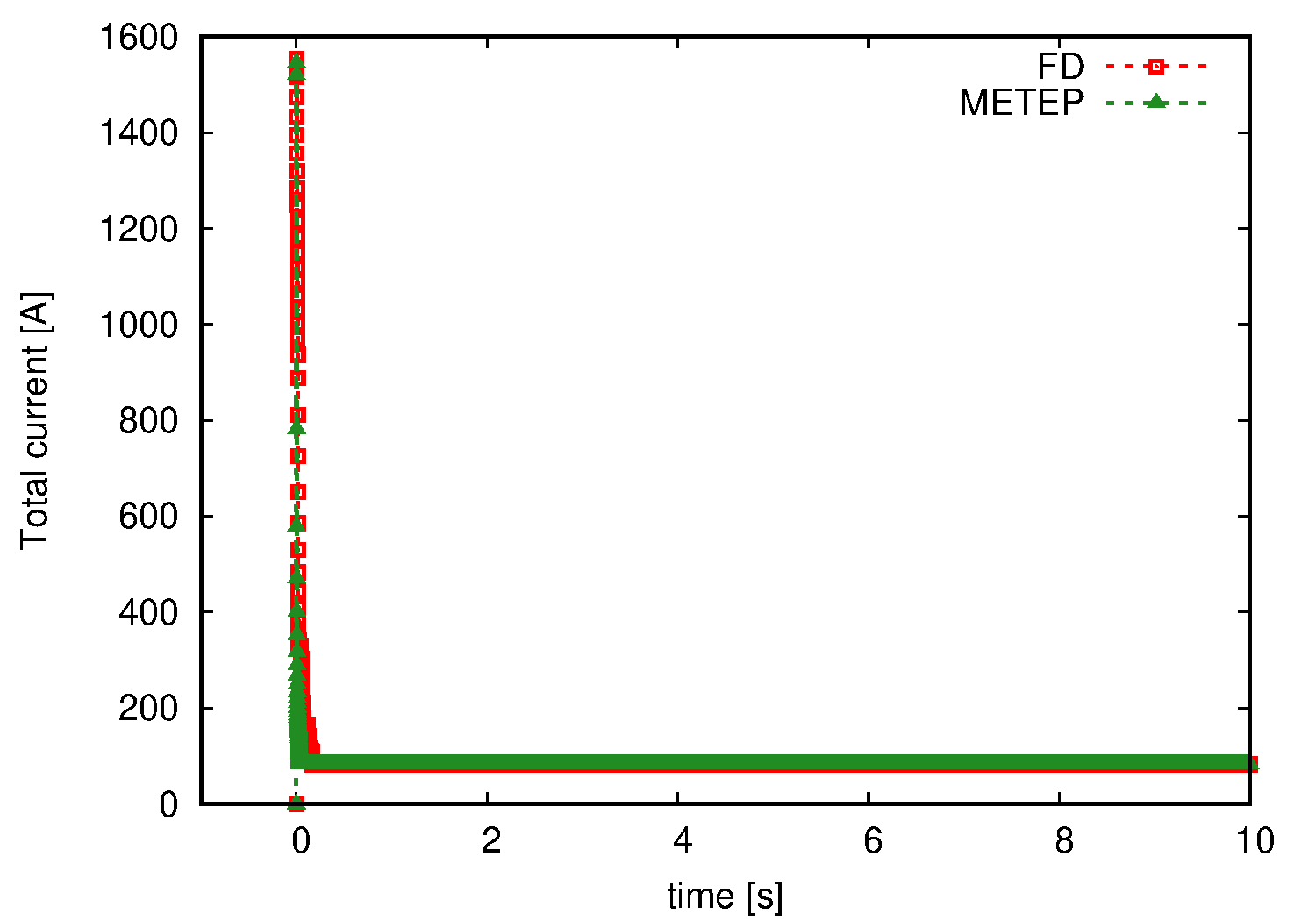}}
\caption{The same as figure \ref{fTlowV} but for an input voltage of 5 V/m.}
\label{fThighV}
\end{figure}

Here, we benchmark the electro-thermal modeling results from METEP and finite differences for two DC input voltages per unit tape length, 0.01 and 5 V/m. These voltages are relatively large for the coil, which causes the current, $I$, to overcome the critical current ($I_c$=150~A or 1200~A$\times$turns at 77 K) shortly after the voltage is switched on. However, there appear screening currents for $I$ below the critical current (see figure \ref{fJmap} for 0.01 V/m and $t=$0.05 and 0.1 s). This current distribution is typical for pancake coils \cite{pardoE2012SSTb}. Above the critical current, the current density is roughly uniform. The temperature distribution is also roughly uniform for all computed cases (see figure \ref{fTmap}).

For 0.01 V/m, the current sharply increases at the beginning, soon overcoming $I_c$ and causing high power dissipation and temperature increase. After reaching a peak of 160 A, or 1280 A$\times$turns, the current slowly decreases (see figure \ref{fTlowV}). The slow decrease in current that follows is due to temperature increase, which reduces $J_c$. Then, $J$ (and $I$) decreases for the same $E$, being the latter dominated by the applied voltage ($E\approx 0.01$ V/m).

For 5 V/m, the coil experiences fast electro-thermal quench, bringing the superconductor above the critical temperature of 92 K and reaching up to 450 K (figure \ref{fThighV}). Although the temperature is high, most REBCO tapes do not get damaged \R{under} these conditions \cite{buranM2019SST, cuninkovaE2022IES}. The fault current (or peak of current) reaches up to 198 A (1584 A$\times$turns), which is 1.32 times the $I_c$ at 77 K. Shortly afterwards, the superconductor overcomes the critical temperature and the current sharply decreases to below 100 A$\times$turns. The low current is caused by the relatively large average resistivity of the metal parts, the stabilization layer \R{being} very thin (2 $\mu$m).

The temperature distribution (figure \ref{fTmap}) and the time evolution of the average temperature and total current (in A$\times$turns) from both models agree (figures \ref{fTlowV} and \ref{fThighV}), mutually confirming the validity of the models. In addition, the temperature at stationary state, $T_\infty$, and the charasteristic time, $\tau$, for $V=5$ V/m agree with the analytical limits of section \ref{s.analytical}, obtained taking the high-temperature limit of $h$ ($h=100$ Wm$^{-2}$K$^{-1}$) \R{and $C_{ve}=2480$ kJ/Km$^3$, $\rho_n=3.17\cdot 10^{-7}$ $\Omega\cdot$m}.

\R{The main advantage of METEP over finite differences (FD) is its reliability because METEP does not require a stability condition, and hence it can take large time steps into account. In addition, METEP can directly solve the steady-state configuration (it is sufficient to use $\Delta t=\infty$). Moreover, METEP can in principle use any element shape, while FD requires rectangual mesh in the coil cross-section. The advantage of FD is that it is faster. In the present implementation, METEP is around 10 times slower than FD, mainly because the minimization routine for METEP has not been optimized. On-purpose minimization algorithms for METEP and parallel computing routines using sectors, such as those used in \cite{pardoE2017JCP}, could reduce the computing time by orders of magnitude. 
}


\section{Conclusion}

This article presented a novel variational method for electro-thermal modelling of superconductors and we benchmarked it to a finite difference method that we implemented coupled with an electromagnetic model. The results agree with each other, as well as with analytical limits. This confirms the correctness of both methods. In addition, we extended the electro-magnetic method MEMEP in order to take input voltage into account, enabling to model screening currents under these conditions.

With these methods, we modeled the electro-thermal behavior of a racetrack coil in short circuit caused by a sudden appearance of a DC voltage. For the studied configuration, the coil could withstand 5 V/m of DC voltage for long times. In addition, the temperature rise is relatively slow, taking around 5 s to \R{reach} 200 K, which is \R{slow} enough for quench detection and switching off the voltage.

The modeling methods presented here could be useful for the design of magnet and power applications, such as high-field magnets or superconducting motors and generators.


\section*{Acknowledgments}

This project has received funding from the European Union's Horizon 2020 research and innovation programme under grant agreement No 951714, and the Slovak Republic from projects APVV-19-0536 and VEGA 2/0036/21. Any dissemination of results reflects only the author's view and the European Commission is not responsible for any use that may be made of the information it contains.

\bibliographystyle{unsrt}


\begin{thebibliography}{10}

\bibitem{hahnS2019Nat}
Seungyong Hahn, Kwanglok Kim, Kwangmin Kim, Xinbo Hu, Thomas Painter, Iain
  Dixon, Seokho Kim, Kabindra~R Bhattarai, So~Noguchi, Jan Jaroszynski, et~al.
\newblock 45.5-tesla direct-current magnetic field generated with a
  high-temperature superconducting magnet.
\newblock {\em Nature}, 570(7762):496--499, 2019.

\bibitem{baiH2020IESa}
Hongyu Bai, Mark~D Bird, Lance~D Cooley, Iain~R Dixon, Kwang~Lok Kim, David~C
  Larbalestier, William~S Marshall, Ulf~P Trociewitz, Hubertus~W Weijers,
  Dmytro~V Abraimov, et~al.
\newblock The {40 T} superconducting magnet project at the national high
  magnetic field laboratory.
\newblock {\em IEEE Trans. Appl. Supercond.}, 30(4):1--5, 2020.

\bibitem{awaji2S021IES}
Satoshi Awaji, Arnaud Badel, Tatsunori Okada, Kohki Takahashi, Hiroshi
  Miyazaki, Satoshi Hanai, Shigeru Ioka, Shinji Fujita, Shogo Muto, Yasuhiro
  Iijima, et~al.
\newblock {Robust REBCO insert coil for upgrade of 25 T cryogen-free
  superconducting magnet}.
\newblock {\em IEEE Trans. Appl. Supercond.}, 31(5):1--5, 2021.

\bibitem{badelA2021IES}
Arnaud Badel, Tatsunori Okada, Kohki Takahashi, Shinji Fujita, Hiroshi
  Miyazaki, Shigeru Ioka, and Satoshi Awaji.
\newblock {Detection and protection against quench/local thermal runaway for a
  30 T cryogen-free magnet}.
\newblock {\em IEEE Trans. Appl. Supercond.}, 31(5):1--5, 2021.

\bibitem{fazilleauP2018SST}
Philippe Fazilleau, Benjamin Borgnic, Xavier Chaud, Fran{\c{c}}ois Debray,
  Thibault L{\'e}crevisse, and Jung-Bin Song.
\newblock Metal-as-insulation sub-scale prototype tests under a high background
  magnetic field.
\newblock {\em Supercond. Sci. Technol.}, 31(9):095003, 2018.

\bibitem{lecrevisseT2022SSTa}
Thibault L{\'e}crevisse, Xavier Chaud, Philippe Fazilleau, Cl{\'e}ment Genot,
  and Jung-Bin Song.
\newblock Metal-as-insulation hts coils.
\newblock {\em Supercond. Sci. Technol.}, 35(7):074004, 2022.

\bibitem{yoonSK2016SST}
Sangwon Yoon, Jaemin Kim, Kyekun Cheon, Hunju Lee, Seungyong Hahn, and
  Seung-Hyun Moon.
\newblock {26 T 35 mm all-GdBa2Cu3O7--x multi-width no-insulation
  superconducting magnet}.
\newblock {\em Supercond. Sci. Technol.}, 29(4):04LT04, 2016.

\bibitem{bhattaraiKR2020SST}
Kabindra~R Bhattarai, Kwanglok Kim, Kwangmin Kim, Kyle Radcliff, Xinbo Hu,
  Chaemin Im, Thomas Painter, Iain Dixon, David Larbalestier, SangGap Lee,
  et~al.
\newblock {Understanding quench in no-insulation (NI) REBCO magnets through
  experiments and simulations}.
\newblock {\em Supercond. Sci. Technol.}, 33(3):035002, 2020.

\bibitem{shaoL2021Ele}
Liangjun Shao, Xintao Zhang, Yufan Yan, Haoyuan Wang, Huajun Liu, and Timing
  Qu.
\newblock {Design of a 20 T class REBCO insert in a 15 T low temperature
  superconducting magnet}.
\newblock {\em Electronics}, 10(14):1741, 2021.

\bibitem{bergenA2019SST}
A~Bergen, R~Andersen, M~Bauer, H~Boy, M~ter Brake, P~Brutsaert, C~B{\"u}hrer,
  M~Dhall{\'e}, J~Hansen, H~ten Kate, et~al.
\newblock Design and in-field testing of the world's first {ReBCO} rotor for a
  3.6 {MW} wind generator.
\newblock {\em Supercond. Sci. Technol.}, 32(12):125006, 2019.

\bibitem{chanWK2012IES}
Wan~Kan Chan and Justin Schwartz.
\newblock A hierarchical three-dimensional multiscale electro--magneto--thermal
  model of quenching in {REBa}$_2${Cu}$_3${O}$_{7-\delta}$
  coated-conductor-based coils.
\newblock {\em IEEE Trans. Appl. Supercond.}, 22(5):4706010--4706010, 2012.

\bibitem{breschiM2017IES}
Marco Breschi, Lorenzo Cavallucci, Pier~Luigi Ribani, Andrey~Vladimir Gavrilin,
  and Hubertus~W Weijers.
\newblock {Modeling of quench in the coupled HTS insert/LTS outsert magnet
  system of the NHMFL}.
\newblock {\em IEEE Trans. Appl. Supercond.}, 27(5):1--13, 2017.

\bibitem{noguchiS2019IES}
So~Noguchi.
\newblock Electromagnetic, thermal, and mechanical quench simulation of {NI
  REBCO} pancake coils for high magnetic field generation.
\newblock {\em IEEE Trans. Appl. Supercond.}, 29(5):1--7, 2019.

\bibitem{markiewiczWD2019SST}
W~Denis Markiewicz, Thomas Painter, Iain Dixon, and Mark Bird.
\newblock {Quench transient current and quench propagation limit in pancake
  wound REBCO coils as a function of contact resistance, critical current, and
  coil size}.
\newblock {\em Supercond. Sci. Technol.}, 32(10):105010, 2019.

\bibitem{badelA2019SST}
Arnaud Badel, Blandine Rozier, Brahim Ramdane, G{\'e}rard Meunier, and Pascal
  Tixador.
\newblock {Modeling of 'quench' or the occurrence and propagation of
  dissipative zones in REBCO high temperature superconducting coils}.
\newblock {\em Supercond. Sci. Technol.}, 32(9):094001, 2019.

\bibitem{gavrilinAV2021IES}
Andrew~V Gavrilin, Dylan~J Kolb-Bond, Kwang~Lok Kim, Kwangmin Kim, William~S
  Marshall, and Iain~R Dixon.
\newblock Quench and stability modelling of a metal-insulation
  multi-double-pancake high-temperature-superconducting coil.
\newblock {\em IEEE Trans. Appl. Supercond.}, 31(5):1--7, 2021.

\bibitem{niuM2021AMM}
Mengdie Niu, Jing Xia, Huadong Yong, and Youhe Zhou.
\newblock Quench characteristics and mechanical responses during quench
  propagation in rare earth barium copper oxide pancake coils.
\newblock {\em Appl. Math. Mech.}, 42(2):235--250, 2021.

\bibitem{genotC2022IES}
C~Genot, T~L{\'e}crevisse, P~Fazilleau, and P~Tixador.
\newblock Transient behavior of a {REBCO} no-insulation or metal-as-insulation
  multi-pancake coil using a partial element equivalent circuit model.
\newblock {\em IEEE Trans. Appl. Supercond.}, 32(6):1--5, 2022.

\bibitem{royF2008IES}
Fran{\c{c}}ois Roy, Bertrand Dutoit, Francesco Grilli, and Fr{\'e}d{\'e}ric
  Sirois.
\newblock {Magneto-thermal modeling of second-generation HTS for resistive
  fault current limiter design purposes}.
\newblock {\em IEEE Trans. Appl. Supercond.}, 18(1):29--35, 2008.

\bibitem{lacroixC2014SST}
Christian Lacroix and Frederic Sirois.
\newblock {Concept of a current flow diverter for accelerating the normal zone
  propagation velocity in 2G HTS coated conductors}.
\newblock {\em Supercond. Sci. Technol.}, 27(3):035003, 2014.

\bibitem{bonnardCH2016SST}
Charles-Henri Bonnard, Fr{\'e}d{\'e}ric Sirois, Christian Lacroix, and
  Ga{\"e}tan Didier.
\newblock {Multi-scale model of resistive-type superconducting fault current
  limiters based on 2G HTS coated conductors}.
\newblock {\em Supercond. Sci. Technol.}, 30(1):014005, 2016.

\bibitem{rivaN2020SST}
Nicolo' Riva, F~Sirois, C~Lacroix, WTB De~Sousa, Bertrand Dutoit, and F~Grilli.
\newblock Resistivity of {REBCO} tapes in overcritical current regime: impact
  on superconducting fault current limiter modeling.
\newblock {\em Supercond. Sci. Technol.}, 33(11):114008, 2020.

\bibitem{gomoryF2022IES}
F~G{\"o}m{\"o}ry, J~{\v{S}}ouc, and M~Mo{\v{s}}at'.
\newblock Formation of hot spots in coated conductors during static and dynamic
  {DC} loading.
\newblock {\em IEEE Trans. Appl. Supercond.}, 32(4):1--7, 2022.

\bibitem{pardoE2017JCP}
E.~Pardo and M.~Kapolka.
\newblock {3D} computation of non-linear eddy currents: Variational method and
  superconducting cubic bulk.
\newblock {\em J. Comput. Phys.}, 344:339--363, 2017.

\bibitem{pardoE2015SST}
E.~Pardo, J.~{\v Souc}, and L.~{Frolek}.
\newblock Electromagnetic modelling of superconductors with a smooth
  current-voltage relation: variational principle and coils from a few turns to
  large magnets.
\newblock {\em Supercond. Sci. Technol.}, 28:044003, 2015.

\bibitem{powellA2002prp}
Adam Powell.
\newblock Finite difference solution of the heat equation.
\newblock {\em Technical Paper}, 2002.
\newblock
  \url{https://dspace.mit.edu/bitstream/handle/1721.1/35256/22-00JSpring-2002/NR/rdonlyres/Nuclear-Engineering/22-00JIntroduction-to-Modeling-and-SimulationSpring2002/55114EA2-9B81-4FD8-90D5-5F64F21D23D0/0/lecture_16.pdf}.

\bibitem{ghabeliA2021SST}
Asef Ghabeli, Mark Ainslie, Enric Pardo, Lo{\"\i}c Qu{\'e}val, and Ratu
  Mataira.
\newblock Modeling the charging process of a coil by an hts dynamo-type flux
  pump.
\newblock {\em Supercond. Sci. Technol.}, 34(8):084002, 2021.

\bibitem{pardoE2012SSTb}
E.~Pardo, J.~{\v{S}}ouc, and J.~Kov{\'a}{\v{c}}.
\newblock {AC} loss in {ReBCO} pancake coils and stacks of them: modelling and
  measurement.
\newblock {\em Supercond. Sci. Technol.}, 25:035003, 2012.

\bibitem{buranM2019SST}
Marek B{\'u}ran, Michal Vojen{\v{c}}iak, Marek Mo{\v{s}}at', Asef Ghabeli,
  Mykola Solovyov, Marcela Pekar{\v{c}}{\'\i}kov{\'a}, L'ubom{\'\i}r Kopera,
  and Fedor G{\"o}m{\"o}ry.
\newblock Impact of a {REBCO} coated conductor stabilization layer on the fault
  current limiting functionality.
\newblock {\em Supercond. Sci. Technol.}, 32(9):095008, 2019.

\bibitem{cuninkovaE2022IES}
Eva Cuninkov{\'a}, Marcela Pekar{\v{c}}{\'\i}kov{\'a}, M~Mo{\v{s}}at, and
  Michal Skarba.
\newblock {Numerical Simulation of Thermal Stabilization Used in HTS Tapes for
  SCFCL Application}.
\newblock {\em IEEE Trans. Appl. Supercond.}, 32(4):1--5, 2022.

\end{thebibliography}

\end{document}